\documentclass[aps,prb,twocolumn,showpacs,amsmath,amssymb]{revtex4}
\usepackage{dcolumn}
\usepackage{bm}
\usepackage{graphicx}
\usepackage{times}
\usepackage{epstopdf}
\begin{document}

\title{Subsurface impurities and vacancies in a three-dimensional topological insulator}
\author{Annica M. Black-Schaffer}
 \affiliation{Department of Physics and Astronomy, Uppsala University, Box 516, SE-751 20 Uppsala, Sweden}
 \author{Alexander V. Balatsky}
\affiliation{Theoretical Division and Center for Integrated Nanotechnologies, Los Alamos National Laboratory, Los Alamos, New Mexico 87545, USA}
\date{\today}

\begin{abstract}
Using a three-dimensional microscopic lattice model of a strong topological insulator (TI) we study potential impurities and vacancies in surface, subsurface, and bulk positions.
For all impurity locations we find impurity-induced resonance states with energy proportional to the inverse of the impurity strength, although the impurity strength needed for a low-energy resonance state increases with the depth of the impurity.
For strong impurities and vacancies as deep as 15 layers into the material, resonance peaks will appear at and around the Dirac point in the surface energy spectrum, splitting the original Dirac point into two nodes located off-center.
Furthermore, we study vacancy clusters buried deep inside the bulk and find zero-energy resonance states for both single and multiple-site vacancies. Only fully symmetric multiple-site vacancy clusters show resonance states expelled from the bulk gap.
\end{abstract}
\maketitle

%
%
\section{Introduction}
\label{sec:introduction}
Topological insulators (TIs) are a new class of quantum matter, where strong spin-orbit coupling results in a bulk energy gap but gapless metallic surface states.\cite{Hasan10, Qi11} In strong TIs, a topological invariant associated with the bulk band structure guarantees the existence of a single (or odd number) surface state with characteristic linear Dirac energy dispersion, where the electron spin is locked to the momentum.\cite{Fu07} 
The surface state is topologically protected against any time-reversal invariant perturbations. This is intimately connected with the absence of backscattering for nonmagnetic impurities, since a spin-flip is required for 180$^\circ$ backscattering. The lack of backscattering was established theoretically early on within a two-dimensional (2D) continuum model for the surface state\cite{Lee09, Zhou09, Guo10} and later also confirmed in experiments.\cite{Roushan09, Zhang09, Alpichshev10} The same 2D surface continuum model finds that, while a local impurity-induced resonance state exists for a potential impurity, its weight diminish as the energy approaches the Dirac point for unitary scatterers and the Dirac point is left unperturbed.\cite{Biswas10}

Surface-only models, however, ignore the finite bulk gap, thus neglecting bulk-assisted processes. 
Using a microscopic 3D lattice model for a strong TI we recently established that a strong impurity on the surface gives rise to a large resonance peak in the local density of states (LDOS) at and around the Dirac point.\cite{Black-Schaffer11TI} Consequently, the topological protection of the Dirac point is destroyed close to the impurity and it splits into two nodes that move off-center. 
Recent scanning tunneling spectroscopy (STS) results\cite{Teague12} on Bi$_2$Si$_3$ have confirmed the existence of such strong resonance peaks at and around the Dirac point. Other experimental data has also shown how localized bound states at defects\cite{Alpichshev11b} and steps \cite{Alpichshev11} do not agree with results from a purely 2D surface continuum model. 

These recent experiments warrant a close investigation of impurities which might give rise to surface resonance states. In particular, since the surface state extends many layers into the material,\cite{Black-Schaffer11TI} even subsurface impurities might significantly affect the LDOS measured on the surface. 
In the opposite limit, the properties of deep subsurface impurities ought to be closely connected to those of bulk impurities. Both potential impurities\cite{Lu11} and finite sized holes\cite{Shan11} in the bulk of a 3D TI have previously been treated within a continuum theory focusing on in-gap bound states. In the case of a finite sized hole, it constitutes an interior surface and will thus necessarily host a surface state in a TI. As the hole radius shrinks, the surface state is transformed into bound states, which are expelled towards the bulk bands due to the finite hole size.
This is in striking contrast to surface single-site vacancies which produce impurity-bound states at the Dirac point. 

In this article we present a comprehensive microscopic study of impurities positioned all the way from the surface to the bulk. In particular, we address the influence of subsurface impurities on the surface LDOS, how bulk impurities behave on a microscopic scale, and we show how the behavior of impurities in these two opposite limits are intimately connected.
More specifically we find that:
i) Subsurface impurities and vacancies as far as 15 layers into the material create a non-dispersive resonance peak in the surface LDOS. Thus, even deep subsurface impurities will affect the low-energy region of the surface state spectrum and be visible in STS measurements.
ii) The resonance energy $E_{\rm res}$ is always inversely proportional to the impurity strength $U$.
However, for the resonance state to enter the low-energy region, the impurity strength needs to be stronger the deeper down the impurity is buried. 
iii) Both impurities and vacancies in the bulk produce in-gap resonance states, connecting smoothly with the behavior of surface impurities and vacancies. These low-energy states will give rise to non-insulating bulk transport.
iv) Fully symmetric multiple-site vacancy clusters have no in-gap resonance peaks, in agreement with continuum results.\cite{Shan11} However, any small deviation from full symmetry produces low-lying resonance peaks. Any realistic microscopically created hole in a 3D TI will therefore have a resonance peak around $E = 0$, mimicking the results of a single vacancy instead of that of a finite size continuum hole. 
%

The rest of the article is organized as follows. In Sec.~\ref{sec:model} we introduce a general microscopic lattice model for studying defects and vacancies in a strong 3D TI. In Sec.~\ref{sec:resultsA} we discuss the surface LDOS and impurity-induced resonance peaks for surface and subsurface impurities and vacancies. In particular, we focus on the dependence of the resonance energy on layer position and impurity strength.
In Sec.~\ref{sec:resultsB} we discuss multiple-site bulk vacancy clusters. We conclude in Sec.~\ref{sec:conclusion} by summarizing our results and discussing experimental consequences.

%
%
\section{Model}
\label{sec:model}
\begin{figure}[htb]
\includegraphics[scale = 0.65]{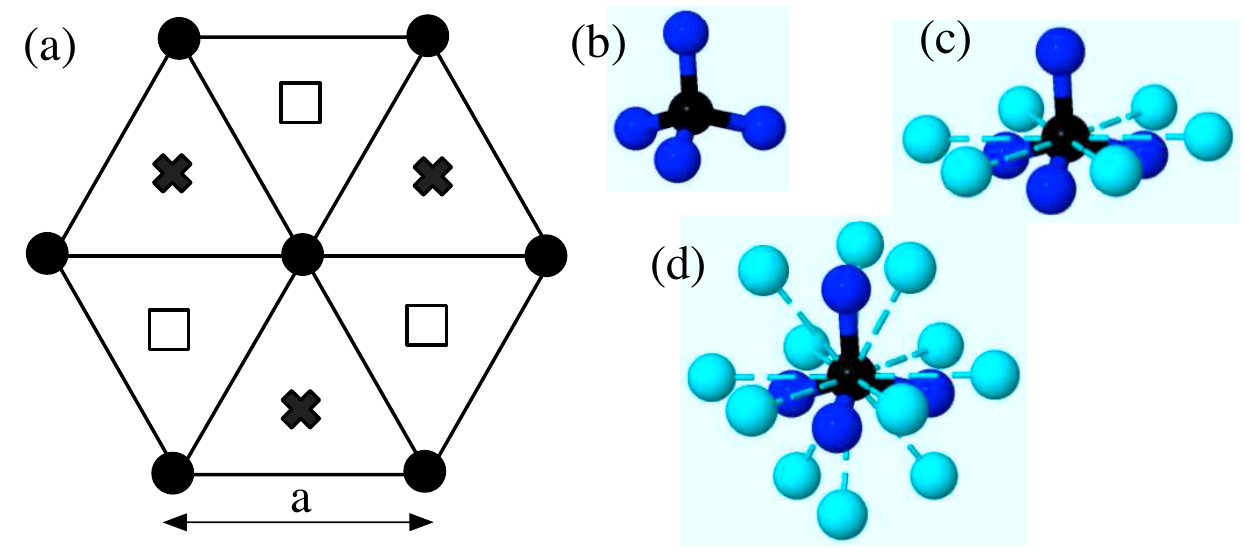}
\caption{\label{fig:lattice} (Color online) (a) Stacking structure for the (111) direction in the diamond lattice. 1st and 2nd A layers (filled circles), 3rd and 4th B layers (crosses), 5th and 6th C layers (squares). In-plane nearest neighbor distance is $a = 1$. Layer separations are $\sqrt{3}a/\sqrt{8}$ for AA layers and $a/\sqrt{24}$ for AB layers. (b) 5-site nearest neighbor cluster with center site (black) and nearest neighbor sites (blue). (c) 11-site nearest neighbor and in-plane next-nearest neighbor cluster with next-nearest neighbor sites (cyan). (d) 17-site next-nearest neighbor cluster.}
\end{figure}

We create a strong TI by using a four band $s$-orbital tight-binding scheme on the diamond lattice with spin-orbit coupling:\cite{Fu07}
%
\begin{align}
\label{eq:H0}
H_0  =  & \ t  \sum_{\langle i,j\rangle} c^\dagger_{i}c_{j} + \mu \sum_i  c^\dagger_i c_j \\ \nonumber
& + \frac{4i\lambda}{a^2}  \sum_{\langle \langle i,j\rangle \rangle} c^\dagger_{i} {\bf s \cdot (d}^1_{ij}\times {\bf d}^2_{ij}) c_{j}.
\end{align}
Here $c_{i}$ is the annihilation operator on site $i$ where we, for simplicity, have suppressed the spin-index. Furthermore, $t$ is the nearest neighbor hopping, $\mu = 0$ the chemical potential, $\lambda = 0.3t$ the next-nearest neighbor spin-orbit coupling, $\sqrt{2}a$ the cubic cell size, ${\bf s}$ the Pauli spin matrices, and ${\bf d}_{ij}^{1,2}$ the two bond vectors connecting next-nearest neighbor sites $i$ and $j$.
By further distorting the hopping amplitude to $1.25t$ along one of the nearest neighbor directions not parallel to the (111) direction, this system becomes a strong TI, with a single surface Dirac cone.\cite{Fu07}
In order to access a surface we create a slab of Eq.~(\ref{eq:H0}) along the (111) direction, see Fig.~\ref{fig:lattice}(a). We are mainly studying slabs with ABBCC...AABBC stacking terminations, hereafter labeled AB termination, but will also compare these results with AABBCC...AABBCC terminated slabs, labeled AA termination, in order to generalize our results.
We choose an energy scale such that the slope of the surface Dirac cone $\hbar v_F\approxeq 1$ for an AB slab, which is achieved by setting $t = 2$ throughout this work.
We find that for slabs with $r \gtrsim 5$ lateral unit cells, where each lateral cell contains six atomic layers, there is only a minimal amount of cross-talk between the two slab surfaces, resulting in a negligible surface energy gap.
We label the different layers in the slab starting with layer 1 for the surface layer. Around layer 15, the remnant DOS of the surface state is becoming negligible and also located close to the bulk gap\cite{Black-Schaffer11TI} and we are thus approaching bulk conditions at this depth.

In order to study the effect of potential impurities we create a rectangular-shaped surface supercell with $n$ sites along each direction. This gives a supercell surface area of $\sqrt{3}n^2a^2/2$ where we use $a=1$ (the nearest neighbor distance on the surface) as the unit of length. We add impurities to our model by adding the term
%
\begin{align}
\label{eq:Himp}
H_{\rm imp} = U \sum_i c_i^\dagger c_i.
\end{align}
to the Hamiltonian in Eq.~(\ref{eq:H0}). Here $U\geq 0$ is the impurity strength and the summation is over all impurity sites in the supercell. We note that by adding $H_{\rm imp}$ we break particle-hole symmetry and thus our model, even with $\mu =0$, corresponds to a rather general situation.

We solve $H= H_0 + H_{\rm imp} = X^\dagger \mathcal{H}X$, where $X^T =(c_{i\uparrow}, c_{i\downarrow})$, 
 in the supercell using exact diagonalization. From the eigenvalues $E^\nu_k$ and eigenvectors $U^\nu_k$ of $\mathcal{H}$, the LDOS resolved at every site $i$ is calculated as
 %
\begin{align}
\label{eq:LDOS}
D_i(E) = \sum_{k,\nu} (|U^\nu_k(i)|^2 + |U^\nu_k(N+i)|^2)\delta(E-E^\nu_k),
\end{align}
where $N$ is the total number of sites and the summation is over all $k$-points in the supercell Brillouin zone and all eigenvalues indexed by $\nu$. The two different terms in Eq.~(\ref{eq:LDOS}) are for spin-up and spin-down electrons, respectively. We will mainly be concerned with the layer-resolved LDOS on nearest neighbor sites to the impurity, which is the average of the site-resolved LDOS in Eq.~(\ref{eq:LDOS}) on nearest neighbor sites in each layer.
We find that a $50 \times 50$ supercell $k$-point grid gives sufficient resolution, while at the same time using a Gaussian broadening of $\sigma = 0.005$ when calculating the LDOS.

%
\section{Results}
\subsection{Subsurface impurities}
\label{sec:resultsA}
We start by studying single-site, isolated, potential impurities with varying impurity strength $U$ including the case of a single vacancy ($U\rightarrow \infty$), which represents the unitary scattering limit. Here we study impurities located from the surface all the way down to the bulk.
\begin{figure}[htb]
\includegraphics[scale = 0.98]{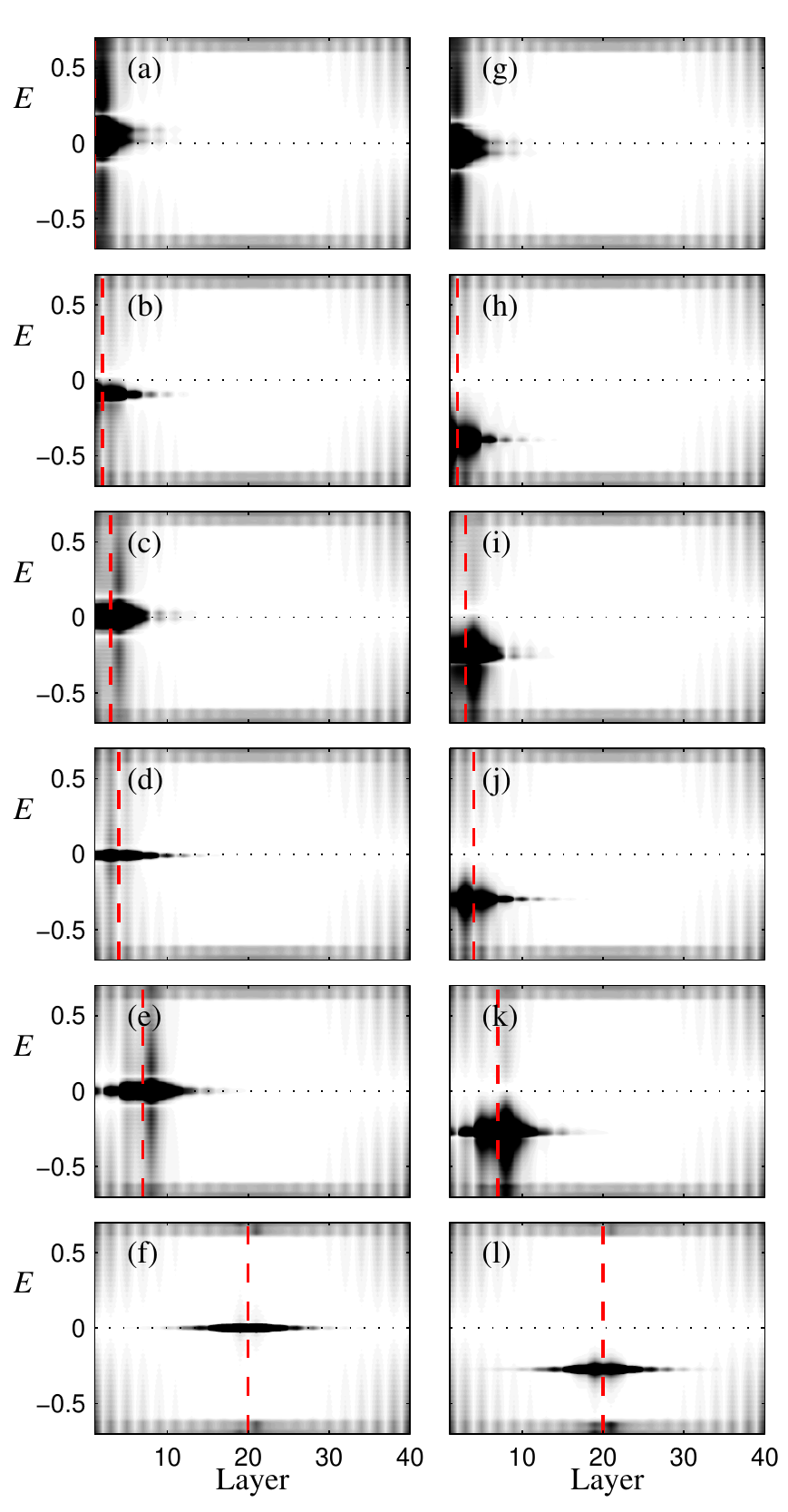}
\caption{\label{fig:LDOSbulk} (Color online) Layer-resolved LDOS averaged over in-plane nearest neighbor sites to a vacancy (a-f) and an $U = 40$ impurity (g-l) positioned in layer 1, 2, 3, 4, 7, 20 (counted from the top) plotted for each layer across a $r =7$ lateral unit cell wide slab with AB termination with a supercell size of $n = 10$. Zero (white), 0.1 (black) states per energy and area unit. Red/Grey dashed vertical lines mark impurity layer, whereas horizontal dotted lines mark $E = 0$.}
\end{figure}
%
Figure~\ref{fig:LDOSbulk} shows nearest neighbor layer-resolved LDOS  for both a vacancy (left column) and a $U = 40$ impurity (right column), positioned in different layers. Starting with a surface layer vacancy (topmost left), there is a wide, double-peak resonance roughly centered at the Dirac point at $E = 0$. As carefully analyzed in Ref.~\onlinecite{Black-Schaffer11TI}, a surface vacancy creates a resonance peak firmly situated on top of the original Dirac point, which splits into two Dirac points situated on either side of the resonance peak. These two Dirac points are the termination points of the valence and conduction Dirac surface states, respectively. 
The local destruction of the topologically protected low-energy Dirac surface state spectrum, and its Dirac point, is due to surface-bulk interaction always present in TIs with a finite bulk band gap.
The width of the resonance peak decreases as the impurity-impurity distance increases with supercell size $n$. However, the total weight of the peak approaches a constant value as $n$ increases,\cite{Black-Schaffer11TI} corroborating the existence of a finite resonance peak even in the limit of a fully isolated impurity. The double-peak structure is also less visible as the impurity-impurity overlap decreases and the center of the peak remains fixed. In fact, the resonance peak is non-dispersive throughout the whole slab for all impurity concentrations and positions.

Since the surface state penetrates relatively deep into the material, by reciprocity argument, impurities positioned in subsurface layers might also have a profound effect on the surface LDOS.
Figures~\ref{fig:LDOSbulk}(b-f) show single vacancies positioned in layer 2, 3, 4, 7, and 20 respectively. There is some oscillation in the energy of the peak as function of layer position, but for all subsurface layer positions $\lesssim 15$, there is still a finite sized resonance peak located at or around $E = 0$ in the surface LDOS. Thus, the original single Dirac point on the surface is destroyed even for vacancies positioned deep into the TI. 
We also find a single-double peak oscillation where double peaks only appear for vacancies in every other layer, but this layer position difference diminishes with increasing impurity-impurity distance.
When approaching the bulk layers, the resonance peak centers firmly at $E = 0$, and its impact on the surface state diminishes as the distance to the surface increases.
In Fig.~\ref{fig:LDOSbulk}(f), the vacancy is positioned deep within the bulk, and there is a narrow, but tall, impurity resonance peak at $E = 0$, but it does not penetrate to the surface.
This result can be understood rather straightforwardly by applying the $T$-matrix formalism to an idealized, but normal, insulator.
In the presence of a scattering potential $\hat{V}$, the Green's function $\hat{G}$ is determined by
%
\begin{align}
\label{eq:G}
\hat{G} = \hat{G}^0 + \hat{G}^0\hat{T}\hat{G}^0,
\end{align}
where $\hat{G}^0$ is the bare Green's function and the $T$-matrix is given by
%
\begin{align}
\label{eq:T}
\hat{T} = (1-\hat{V}\hat{G}^0)^{-1}\hat{U}.
\end{align}
Since the poles of the Green's function give the energy spectrum for single-particle excitations, we can find the energy $E_{\rm res}$ of any impurity-induced resonance state by searching for poles in the $T$-matrix.
For an atomically sharp impurity, described by the $\delta$-function potential $\langle x| \hat{U}|x\rangle = U\delta(x)$, the resonance energy is given by
%
\begin{align}
\label{eq:res}
\frac{1}{U} = {\rm Re} [G^0(E_{\rm res})],
\end{align}
as long as ${\rm Im}[G^0(E_{\rm res})]$ is sufficiently small.\cite{Balatsky06}
Using an idealized insulator with $k$-independent valence and conductions bands separated by a band gap $E_g$, the bare Green's function is $G^0(\omega, {\bf k}) = (\omega - E_g/2 + i\eta)^{-1} + (\omega + E_g/2 - i \eta)^{-1}$, with $\eta$ infinitesimal small. Thus Eq.~(\ref{eq:res}) gives $E_{\rm res} \rightarrow 0$ as $V \rightarrow \infty$.
If, on the other hand, the TI has a finite doping such that the Fermi energy $E_F =  E_D + x \equiv 0$, where $E_D$ is the energy of the Dirac point and $|x|<E_g/2$ for $E_F$ to still be inside the bulk gap, the same argument gives $E_{\rm res} = -x = E_D$. That is, the resonance will always be situated at the Dirac point for a unitary impurity, independent of the doping of the system. We have confirmed this result numerically by including a finite chemical potential in Eq.~(\ref{eq:H0}).
The above derivation is dependent on the valence and conduction bands being mirror-symmetric with respect to $E_D$ for all $k$-values. While this is true in our model TI, it is in general not true in a real material. However, as long as valence and conduction bands are approximately mirror-symmetric in $E_D$ in the part of the Brillouin zone where the band gap is as smallest, we expect our results to still be qualitatively correct. If on the other hand, the energy difference $E_c$ between conduction band and Dirac point, and $E_v$ between valence band and Dirac point are different, the resonance energy is instead $E_{\rm res} = -x + (E_c-E_v)/2$ which is located away from the Dirac point.
This $T$-matrix calculation is also important as it shows that our results are independent of the particular lattice model.

%
The LDOS for a finite $U$-impurity are very similar to those of a single vacancy. The main difference is that the resonance peak in general do not appear at or around $E = 0$, unless $U$ is large, and thus do not destroy the low-energy features of the Dirac surface state. There is also a clear trend that the deeper the impurity, the larger the $U$ needed for a low-energy resonance peak. This is clearly seen in in Figs.~\ref{fig:LDOSbulk}(g-l) where a $U = 40$ surface impurity is seen to destroy the original Dirac point, but where the same impurity in subsurface layers produces an impurity-induced resonance away from $E = 0$.
Figure~\ref{fig:LDOSbulk}(l) shows how a bulk $U = 40$ impurity clearly produces an in-gap resonance peak, associated with a state tightly bound to the impurity site. It was recently argued, based on results from a continuum model, that a non-magnetic $\delta$-function impurity cannot produce in-gap bound states in a 3D TI.\cite{Lu11} Our results, however, show that the closest lattice equivalent of a $\delta$-function, i.e.~the single-site impurity, clearly produces in-gap bound states. This result is true as long as the impurity strength is large enough to put the resonance peak within the bulk gap. In our model system that means $U \gtrsim 20$. For smaller $U$ there is still a resonance state but it is located at energies above the bulk gap.

\begin{figure}[htb]
\includegraphics[scale = 0.98]{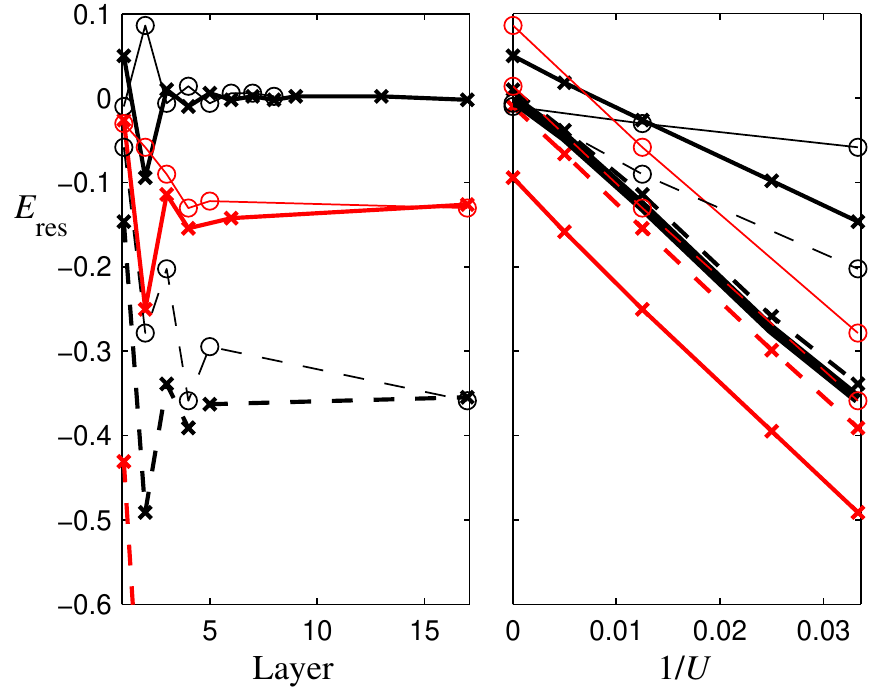}
\caption{\label{fig:peakpos} (Color online) (a) Impurity resonance peak position as function of layer position for AB surface termination (thick lines, $\times$) and AA surface termination (thin lines, $\circ$) for a vacancy ($U = \infty$) (solid black), $U = 80$ (solid red), $U = 30$ (dashed black), and $U = 14$ (dashed red) impurities, where the last set of results are only displayed within the bulk gap $E_g \approx 0.6$.
(b) Impurity resonance peak position as function of the inverse impurity strength $1/U$ for AB surface termination (thick lines, $\times$) and AA surface termination (thin lines, $\circ$) for impurity layer position 1 (solid black), 2 (solid red), 3 (dashed black), 4 (dashed red), and bulk (thickest black).}
\end{figure}

In Fig.~\ref{fig:peakpos} we analyze in more detail the resonance energy peak position $E_{\rm res}$, extracted from the layer-resolved LDOS surface spectra, as function of both impurity layer position (a) and impurity strength (b). Since the resonance peak is non-dispersive, the peak energy position is the same in all layers.
As clearly seen in Fig.~\ref{fig:peakpos}(a), the resonance peak appear at larger (negative) energies, i.e.~farther from the low-energy region, for subsurface impurity positions. Thus for an impurity to influence the low-energy region of the surface Dirac spectrum it needs to be stronger the farther it is from the surface. It is also clear that the resonance peak move toward the low-energy region from larger (negative) energies as $U$ increases. This is equally true for both surface and subsurface positions.
Apart from these trends, there is also a layer oscillation in the peak position but it quickly dies out as the impurity position approaches the bulk.
We have here also included results for an AA terminated surface (thin lines, $\circ$) alongside the AB surface results (thick lines, $\times$). We note that the specifics of the layer oscillations are somewhat surface dependent as the AA surface termination produces slightly different results, but, in general, both surface terminations display remarkably similar results.
In Fig.~\ref{fig:peakpos}(b) we plot the peak position as function of the inverse impurity strength $1/U$. For all impurity layer positions, including both the surface and the bulk, the peak position is proportional to $1/U$, with $E_{\rm res} = k/U + m$.
For AB surface termination, the slope $k$ is approximately constant between different impurity layer positions but the off-set $m$ varies for impurities close to the surface. For AA surface termination there is also some variation of the slope $k$ between layer positions. However, already for impurities in layer 4, the peak position is largely set by the bulk behavior (thickest line). The $1/U$-dependence for the resonance peak position in the bulk follow directly from the same $T$-matrix argument given above and the $1/U$-dependence for surface impurities has been established using a 2D continuum model for the surface state.\cite{Biswas10} However, the resonance peak was in the latter case found to disappear at unitary scattering, something we most notably do not see in our microscopic lattice model.
To summarize this section, we conclude that subsurface and bulk impurities, behave very similar to surface impurities, with a $1/U$ resonance peak energy dependence, although a stronger impurity is needed in subsurface positions in order to observe in-gap resonances. The non-dispersiveness of the resonance peak means that for any finite impurity-surface coupling, a resonance peak will also be present in the surface energy spectrum. We find that resonance peak traces are clearly present in the surface LDOS for impurities as far down as $\sim 15$ layers below the surface. Moreover, both finite strength impurities and vacancies in the bulk produce low-energy resonance states, a result which connects smoothly with the behavior of impurities close to the surface.

\subsection{Bulk vacancy clusters}
\label{sec:resultsB}
The $E = 0$ resonance peak present for a single-site bulk vacancy is associated with a very tightly bound state around the vacancy site. As Eq.~(\ref{eq:res}) showed, the $E_{\rm res}=0$ peak is the same as that of a vacancy in an idealized normal insulator, and is thus a very robust result.
On the other hand, Shan {\it et al.}\cite{Shan11}~recently used a continuum model to demonstrate the existence of bound states for a finite sized hole in a TI. In that case the bound states are simply a manifestation of the fact that a finite sized hole creates an interior surface in the TI. Holes with a very large radius $R$ possess a surface state very similar to that of a planar surface, although, technically, the surface state will have to obey periodic boundary conditions around the hole. As the radius $R$ becomes finite, the surface state turns into bound states with an energy separation which gets larger with decreasing $R$. Finally, for small enough holes the bound states are expelled to the bulk bands. Most notably, this continuum model do {\it not} produce $E = 0$ bound states for any size holes, unless $R \rightarrow \infty$. Clearly this result is at odds with our microscopic result for a single-site vacancy. To further shed light on this discrepancy we have studied highly-symmetric bulk vacancy clusters, involving as many as 17 sites, in order to increase the effective radius of our microscopically created hole.
%
\begin{figure}[htb]
\includegraphics[scale = 0.98]{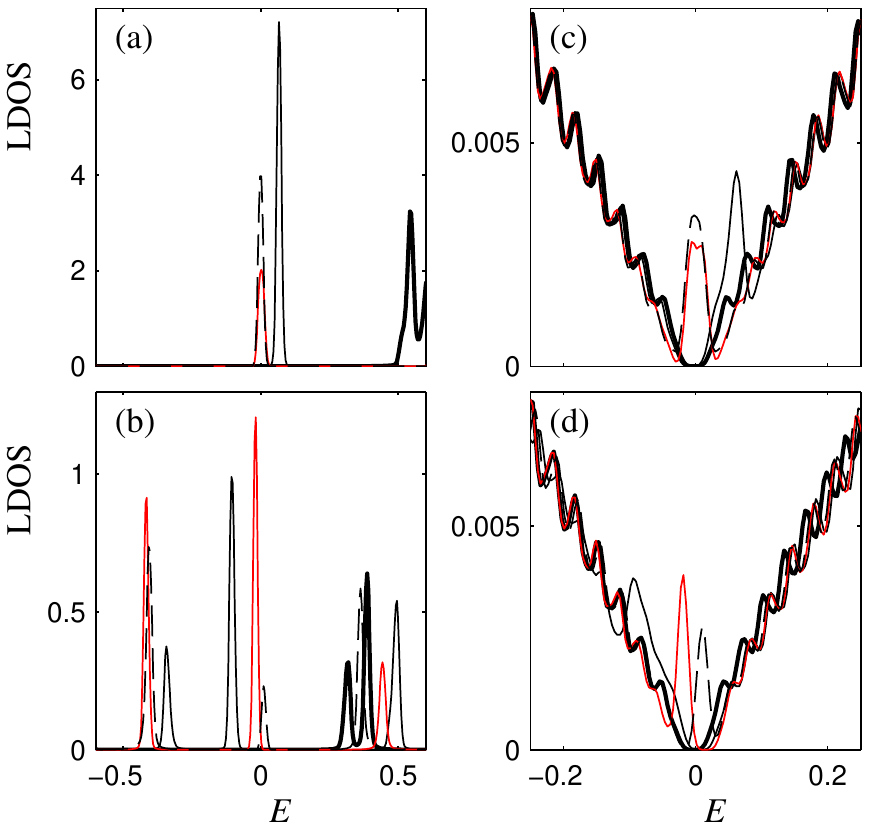}
\caption{\label{fig:vacclusters} (Color online) LDOS averaged over in-plane nearest neighbor sites in layer 17 (a, b) and layer 1 (c, d) for different highly symmetric vacancy clusters centered at layer 17. (a, c): 5-site nearest neighbor vacancy cluster with radius $\sqrt{3}a/\sqrt{8}$ (thick black), 5-site nearest neighbor cluster with 1 next-nearest neighbor substitution (thin black), and 2 next-nearest neighbor substitutions (red), 1-site single impurity (dashed). (b, d): 17-site next-nearest neighbor vacancy cluster with radius $a$ (thick black), 17-site next-nearest neighbor cluster with two different 1 next-next-nearest neighbor substitutions (thin black and red), 11-site cluster consisting of the 4 nearest neighbors and the 6 in-plane next-nearest neighbors (dashed). Small finite gap at $E = 0$ in the surface state is due to the finite width of the slab ($r = 6$). Surface termination is AB and the supercell size is $n = 10$.}
\end{figure}
%
Figure~\ref{fig:vacclusters}(a) shows the LDOS on nearest neighbor sites to both a single-site vacancy (dashed line) and three different 5-site vacancy clusters. The diamond lattice has four nearest neighbors situated at the corners of a tetrahedron, a distance $\sqrt{3}a/\sqrt{8} \approx 0.6a$ from the center site, see Fig.~\ref{fig:lattice}(b). For such a 5-site nearest neighbor vacancy cluster, the resonance peak move up close to the bulk band gap at $E_g \approx 0.6$ (thick line). However, if we replace one of the nearest neighbors with a next-nearest neighbor, the impurity-bound state reappears close to $E = 0$ (black line). Further distortion by replacing two nearest neighbors with next-nearest neighbors creates a resonance state at $E = 0$ (red/grey line), the same result as for the single-site vacancy. We see in Fig.~\ref{fig:vacclusters}(c) how these peaks also show up as extremely small impurity resonances in the surface LDOS at the same energies when these vacancies are centered around layer 17.
In Fig.~\ref{fig:vacclusters}(b, d) we show the same result for even larger vacancy clusters. The diamond (111) slab has six in-plane next-nearest neighbors and an additional six next-nearest neighbors out-of-plane, situated a distance $a$ from the center site, see Figs.~\ref{fig:lattice}(c,d). An 11-site cluster including the four nearest neighbors and the six in-plane next-nearest neighbors creates a resonance around $E = 0$ (dashed line). When including all next-nearest neighbors into a fully-symmetric 17-site cluster, the resonance peaks move up to around $E = 0.4$ (thick line). However, distorting this 17-site cluster by exchanging only one next-nearest neighbor for a next-next-nearest neighbor again produces peaks in the very low-energy part of the spectrum (black and red/grey lines).
We thus find that fully-symmetric vacancy clusters involving all nearest and next-nearest neighbor sites expels the impurity-bound states to high energies, in accordance with earlier continuum model results. Also, when increasing the radius from $0.6a$ for the nearest neighbor cluster to $a$ for the next-nearest neighbor cluster, the resonance peak moves to slightly lower energies, in agreement with the continuum results. However, even the smallest possible distortion of either of these two clusters produces results more resembling  those of a single-site vacancy, where the resonance peak sits firmly at $E  = 0$. Thus, despite the topological origin of the surface state in a TI, there is a surprisingly large sensitivity to small deviations in the cluster shape.
Since any microscopically sized hole in a TI will likely have some asymmetry, we conclude that even for fairly large such holes, the continuum limit will not be reached, but a resonance peak will be present at or around $E = 0$.

%
\section{Concluding remarks}
\label{sec:conclusion}
Using a 3D microscopic lattice model of a strong TI we have shown that strong potential impurities and vacancies create low-lying impurity-bound resonance peaks, with an $1/U$-dependence for the resonance peak energy for impurities in any layer, including the bulk. Impurities as far as 15 layers below the surface have resonance peaks visible in the surface LDOS. This is also approximately the penetration depth of the surface state into the interior of the TI. Thus any vacancy or unitary impurity, within the penetration depth of the TI surface state, produces a peak in the LDOS at or very near the Dirac point, which is subsequently destroyed and split into two nodes that move off-center.
Recent STS data\cite{Teague12} on nonmagnetic unitary impurities in Bi$_2$Se$_3$ has shown sharp energy resonance peaks at the Dirac point, with diverging strength as the Fermi level approaches the Dirac point.
Our results show that the impurities do not necessarily have to be located on the surface, but also subsurface impurities can generate such surface resonance peaks.
The experimental presence of strong resonance states at the Dirac point confirms the need for a 3D model, which explicitly includes bulk states, since 2D continuum results do not find any strong resonance peaks near the Dirac point.\cite{Biswas10} 
For surface impurities the resonance peak decays quickly, approximately as $1/R^3$ with distance on the surface,\cite{Black-Schaffer11TI} and we find a similar dependence for subsurface impurities in the surface LDOS. This fast decay should be contrasted with a rather extended spread perpendicular to the surface.
Experimentally, the resonance peaks were found to decay within as little as 2\AA, which is in qualitative agreement with our results. Such fast decay signals a quick healing of the single Dirac point spectrum, as would be expected for a topologically protected surface.
The impurity-induced resonance peaks in the surface state in a TI are, in fact, similar to impurity resonances in graphene \cite{Ugeda10} and $d$-wave high-temperature superconductors,\cite{Balatsky06} two other materials with Dirac-like low-energy spectra. Thus, once any topological protection is lost due to strong scattering, there is a strong argument for a unified local response to impurities for all ``Dirac" materials.\cite{Wehling11} This unified response corroborates the model-independence of our numerical results, as it is only the Dirac-like surface state, in combination with a finite bulk gap, that is important.

Closely connected to the behavior of near-surface impurities are that of bulk impurities, where we find $E_{\rm res} = 0$ peaks for single-site vacancies in the bulk. This result does not agree with continuum model results for finite holes in a TI.\cite{Shan11} To expand on this discrepancy we have studied extended bulk vacancy clusters. We find that, while fully-symmetric 5- and 17-site clusters do not have any low-energy resonance states in agreement with continuum results, any asymmetry in the clusters produces $E_{\rm res} \approx 0$ resonance peaks. Since any vacancy cluster of microscopic origin is likely to not be fully symmetric, we conclude that a microscopic approach is required for such holes.
For a finite strength bulk impurity, we similarly find contradictions with continuum model results. The $1/U$-dependence for the resonance energy produces in-gap resonances for strong impurities, in contrast to the absence of in-gap states for $\delta$-potential impurities in continuum models. \cite{Lu11}
In fact, both the bulk vacancy $E_{\rm res} = 0$ resonance state and the $1/U$ bulk impurity energy dependence are independent of the topological index and also present in a trivial band insulator, as we show by a simple T-matrix calculation. As a consequence, these conclusions for bulk impurities are independent on the specifics of the lattice model. 
%
The low-energy resonance peaks for deep subsurface and bulk impurities can have a profound effect on the conductivity, as they can give rise to gapless bulk conductivity, thus masking the surface transport properties. Moreover, in the presence of a finite overlap between surface and bulk vacancy states, the surface electrons can be scattered by these zero-energy resonance states. In the limit of dense vacancy concentration, vacancy-band formation will allow edge-edge transitions, thus opening a gap in the topologically protected surface state.

%
\begin{acknowledgments}
We are grateful to R.~Biswas, Z.~Hasan, D.-H.~Lee,  H.~Manoharan, N.~Nagaosa, A.~Wray, S.-C.~Zhang for discussions. AMBS acknowledges support from the Swedish research council (VR). Work at Los Alamos was supported by US DoE Basic Energy Sciences and in part by the Center for Integrated Nanotechnologies, operated by LANS, LLC, for the National Nuclear Security Administration of the U.S. Department of Energy under contract DE-AC52-06NA25396 and by UCOP.
\end{acknowledgments}


\end{document}